\documentclass[floatfix,prb,aps,a4paper,twocolumn]{revtex4}

\usepackage{pgf}
\usepackage{pgffor}

 \usepackage{bm}
 \usepackage{amsmath}
 \usepackage{amssymb}
 \usepackage{latexsym}
 \usepackage{amsfonts}
 \usepackage{epsfig}
 \usepackage{subfigure}
 \usepackage{color}
 \definecolor{darkblue}{rgb}{0,0,.5}
 \usepackage[linktocpage, colorlinks=true ,linkcolor=darkblue, citecolor=darkblue]{hyperref}
 \usepackage[all]{hypcap}

 \newcommand{\expval}[1]{\left< #1 \right>}

 \newcommand{\nn}{\nonumber\\}
 
 \newcommand{\f}[1]{\mbox{\boldmath$#1$}}

 \newcommand{\bea}{\begin{eqnarray}}
 \newcommand{\ea}{\end{eqnarray}}
 \newcommand{\eea}{\end{eqnarray}}
 
 \newcommand{\trace}[1]{{\rm Tr}\left\{ #1 \right\}}

 \newcommand{\abs}[1]{{\left| #1 \right|}}

\begin{document}

\title{Probing the power of an electronic Maxwell Demon}

\author{Gernot Schaller}
\email{gernot.schaller@tu-berlin.de}
\author{Clive Emary}
\author{Gerold Kiesslich}
\author{Tobias Brandes}

\affiliation{Institut f\"ur Theoretische Physik, Technische Universit\"at Berlin, Hardenbergstr. 36, 10623 Berlin, Germany}

\begin{abstract}
We suggest that a single-electron transistor continuously monitored by a quantum point contact may function as Maxwell's demon
when closed-loop feedback operations are applied as time-dependent modifications of the tunneling rates across its junctions.
The device may induce a current across the single-electron transistor even when no bias voltage or thermal gradient is applied.
For different feedback schemes, we derive effective master equations and compare the induced 
feedback current and its fluctuations as well as the generated power.
Provided that tunneling rates can be modified without changing the transistor level, the device may be implemented with 
current technology.
\end{abstract}

\maketitle


Maxwell's demon -- a hypothetical intelligence in a box capable of sorting hot (fast) and cold (slow) atoms to left and right sub-cavities 
simply by matchingly inserting and removing an impenetrable wall -- was
initially suggested by J. C. Maxwell to highlight that thermodynamics is a macroscopic theory.
Under ideal conditions, inserting and removing the wall would not require work and the 
apparent contradiction with the second law -- the entropy in the box would be reduced and after the sorting process, 
work could be extracted from the temperature difference between the sub-cavities -- has been a source of much debate ever since~\cite{maruyama2009a}.
It is now generally believed that this paradox is overcome by the Landauer principle~\cite{landauer1961a}:
Deleting the data required for the processing in the demons mind would at least generate entropy $S=k_B \ln 2$ or dissipate heat of at least 
$Q= k_B T \ln 2$ per bit of information, yielding a net production of entropy.

\begin{figure}[ht]
\includegraphics[width=0.48\textwidth]{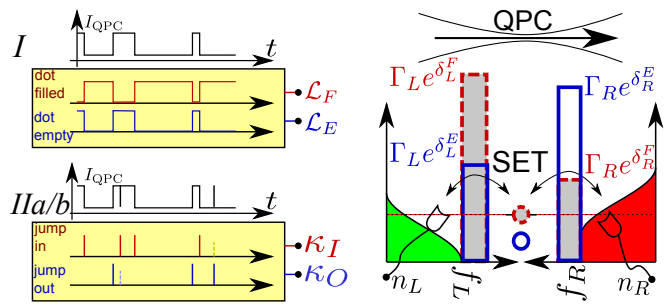}
\caption{\label{Fsketch}(Color Online) 
Sketch of the setup: An SET with two attached leads (sketched are their Fermi functions $f_L$ and $f_R$) is capacitively coupled
to a nearby QPC.
Its current $I_{\rm QPC}$ yields information on the instantaneous occupation of the SET.
One may either apply Liouvillians ${\cal L}_E$ and ${\cal L}_F$ with different tunneling rates conditioned on the {\em present state of the SET} 
(scheme I, compare different tunneling barriers in SET sketch) 
or control operations $\kappa_I$ and $\kappa_O$ conditioned on the {\em change of the SET state} (schemes IIa/b, not shown in SET sketch).
During the control operations, electrons may also tunnel into or out of the SET (dashed shorter spikes), which may not (IIa) or may (IIb)
recursively trigger further control operations.
}
\end{figure}

The demon performs a measurement on the system (is the atom slow or fast) and conditioned on the measurement result
it performs an action (opening or closing the wall), which is formally nothing but a closed-loop feedback control scheme.
In our proposal, we replace the two sub-cavities of the box by two conductors that we assume to be in separate thermal equilibria.
The conductors are coupled by a single resonant level, which does not require charging and spin~\cite{datta2008a} effects.
Conditioned on its occupation dynamics -- provided by a nearby 
quantum point contact (QPC) -- the tunneling rates to the conductors are modified in time.
The feedback schemes we consider here are illustrated in Fig.~\ref{Fsketch}.
We will address the power production, current and statistical properties of the device.

\section{Method}\label{Smethod}
A single-electron transistor (SET) coupled to two leads in thermal equilibria may in sequential tunneling (weak coupling)
approximation be well described by a Markovian Lindblad-type master equation $\rho(t) = {\cal L} \rho(t)$, where 
-- when written as a vector --
$\rho(t)=(\rho_0(t),\rho_1(t))$ describes the SET part with occupation probability $\rho_1(t)$.
The Liouvillian ${\cal L}$ can be derived from microscopic calculations or simply using Fermis golden rule and
reads~\cite{esposito2009b}
\bea\label{Emaster}
{\cal L} = \sum\limits_{\alpha\in\{L,R\}} \Gamma_\alpha
\left(\begin{array}{cc}
-f_\alpha & +(1-f_\alpha)\\
+f_\alpha & -(1-f_\alpha)
\end{array}\right)\,,
\eea
where $\Gamma_\alpha$ denotes the tunneling rate between SET and lead $\alpha$ and 
$f_\alpha\equiv\left[e^{\beta_\alpha(\epsilon-\mu_\alpha)}+1\right]^{-1}$ the Fermi function of lead $\alpha$ with
corresponding inverse temperature $\beta_\alpha$ and chemical potential $\mu_\alpha$ at the
SET level $\epsilon$ (assumed to be constant throughout).
Since the total number of electrons is conserved in the tunneling processes, it is possible to uniquely identify matrix 
elements of the Liouvillian with electronic jump processes into and out of the left
and right leads, which enables one to convert Eq.~(\ref{Emaster}) into an infinite set of coupled equations for $(n_L,n_R)$-resolved
density matrices $\rho^{(n_L,n_R)}(t)$.
These are conditioned on the number of electrons that have tunneled after time $t$ into or out of the left ($n_L$) and
right ($n_R$) leads.
Fourier-transformation, 
\mbox{$\rho(\chi_L,\chi_R) = \sum_{n_L,n_R} \rho^{(n_L,n_R)}(t) e^{i\chi_L n_L} e^{i\chi_R n_R}$}, yields the compact equation
$\dot{\rho}(\chi_L,\chi_R,t) = {\cal L}(\chi_L,\chi_R) \rho(\chi_L,\chi_R,t)$ with
\bea\label{Emastern}
{\cal L}(\chi_L,\chi_R) &=& \sum\limits_{\alpha\in\{L,R\}} \Gamma_\alpha
\left(\begin{array}{cc}
-f_\alpha & +(1-f_\alpha) e^{+i\chi_\alpha}\\
+f_\alpha e^{-i\chi_\alpha} & -(1-f_\alpha)
\end{array}\right)\nn
&\equiv&  \sum\limits_{\alpha\in\{L,R\}} \Gamma_\alpha {\cal F}_\alpha(\chi_\alpha)\,.
\eea
In absence of feedback, the above equation yields the complete statistics of electrons
(current, noise, skewness etc.) that have tunneled across the SET~\cite{nazarov2003}.
At small SET bias, the QPC does not resolve to which of the attached leads a tunneling process has happened.
Therefore, simplest feedback operations can only be conditioned on whether the SET is empty or filled (scheme I) or whether an electron has
jumped into or out of the SET (schemes IIa/b).
Such control operations may easily be included in the numerical solution of Eq.~(\ref{Emastern}) via an 
associated stochastic Schr\"odinger equation~\cite{breuer2002}.
For analytic results however, it is more favorable to derive an effective Liouvillian under feedback 
(compare appendices~\ref{AeffmasterI},~\ref{AeffmasterIIa}, and~\ref{AeffmasterIIb}), which 
enables one to systematically study the effects of feedback on the statistics.

In {\em scheme I}, we only apply two different Liouvillians -- conditioned on whether the SET is empty 
(${\cal L}_E(\chi_L,\chi_R)\equiv\sum_{\alpha\in\{L,R\}} \Gamma_\alpha e^{\delta_\alpha^E} {\cal F}_\alpha(\chi_\alpha)$)
or filled 
(${\cal L}_F(\chi_L,\chi_R)\equiv\sum_{\alpha\in\{L,R\}} \Gamma_\alpha e^{\delta_\alpha^F} {\cal F}_\alpha(\chi_\alpha)$), 
where the dimensionless feedback parameters $\delta_\alpha^{E/F} \in \mathbb{R}$ encode the modification of tunneling rates
and $\delta_\alpha^{E/F}=0$ recovers the situation without feedback.
The effective feedback generator reads in this case 
\bea\label{EfbI}
{\cal L}_{\rm fb}^I(\chi_L,\chi_R) &=& {\cal L}_E (\chi_L,\chi_R)
\left(\begin{array}{cc}
1 & 0\\
0 & 0
\end{array}\right)\nn
&&+{\cal L}_F(\chi_L,\chi_R)
\left(\begin{array}{cc}
0 & 0\\
0 & 1
\end{array}\right)\,.
\eea
An extremal form of this feedback scheme would be to cut the left junction as soon as the SET is filled ($\delta^F_L\to-\infty$) and to cut the right
junction when it is empty ($\delta^E_R\to-\infty$), which automatically implies unidirectional transport -- independent of potential or temperature gradients.
This effectively implements a feedback ratchet with two teeth -- compare SET tunneling barriers in Fig.~\ref{Fsketch}.
However, in the idealized classical limit where the electrons are localized
either on the left lead, the SET, or the right lead, this ratchet does not directly perform work as its potential does not change where the 
electron is localized.
The resulting effective Liouvillian ${\cal L}_{\rm fb}^I$ cannot generally be written in the form of Eq.~(\ref{Emaster}) using modified tunneling rates $\tilde\Gamma_\alpha>0$
and Fermi functions $\tilde f_\alpha \in (0,1)$.

In {\em scheme IIa}, we instantaneously modify the SET tunneling rates by $\delta$-kicks (compare appendix~\ref{Adeltakick}) immediately after an electron has jumped in
($e^{\kappa_I}$) or out ($e^{\kappa_O}$) of the SET, such that
\bea
\kappa_I(\chi_L,\chi_R) &=& \sum_{\alpha\in\{L,R\}} \delta^I_\alpha {\cal F}_\alpha(\chi_\alpha)\,,\nn
\kappa_O(\chi_L,\chi_R) &=& \sum_{\alpha\in\{L,R\}} \delta^O_\alpha {\cal F}_\alpha(\chi_\alpha)\,,
\eea
where $\delta^{I/O}_{R/L}\ge 0$ are dimensionless parameters roughly given by the product of pulse width and height of time-dependent SET tunneling rates (compare appendix~\ref{Adeltakick}).
The counting-field dependence arises from the simple fact that during the control operation
electrons may tunnel through the junctions (in scheme IIa, these tunneling events do not trigger further control operations):
In fact, for infinitely strong feedback at one junction and zero feedback at the other junction (e.g., $\delta^I_L\to\infty$ and $\delta^I_R\to 0$) one obtains immediate equilibration of
the SET with the lead to which it is tunnel-coupled (e.g., $e^{\kappa_I(0,0)}\rho \to (1-f_L,f_L)$ for all $\rho$).
These tunneling events have to be taken into account when the complete statistics is required.
Using from Eq.~(\ref{Emastern}) the decomposition
\bea
{\cal F}_\alpha(\chi_\alpha)\equiv{\cal F}_\alpha^0 + {\cal F}_\alpha^+ e^{+i\chi_\alpha} + {\cal F}_\alpha^- e^{-i\chi_\alpha}\,,
\eea
where ${\cal F}_\alpha^+$ (${\cal F}_\alpha^-$) is responsible for jumps into (out of) contact $\alpha$ and ${\cal F}_\alpha^0$ conserves the system
occupation, the full effective feedback Liouvillian reads
\bea\label{EfbIIa}
{\cal L}_{\rm fb}^{IIa}(\chi_L,\chi_R) &=& \sum_{\alpha\in\{L,R\}} \Gamma_\alpha {\cal F}_\alpha^0\nn
&&+ e^{\kappa_O(\chi_L,\chi_R)}  \sum_{\alpha\in\{L,R\}} \Gamma_\alpha {\cal F}_\alpha^+ e^{+i \chi_\alpha}\nn
&&+ e^{\kappa_I(\chi_L,\chi_R)} \sum_{\alpha\in\{L,R\}} \Gamma_\alpha {\cal F}_\alpha^- e^{-i \chi_\alpha}\,,
\eea
such that jumps in and out of the system are immediately followed by the respective control operation.
In contrast to feedback schemes modifying the system Hamiltonian~\cite{wiseman2010,kiesslich2011} it is 
no longer possible to simply shift counting fields by control operators.

For an infinitely fast QPC sampling rate, it would also be possible to recursively trigger further control operations when
tunneling events take place during control.
In {\em scheme IIb} we restrict ourselves to the case that during the control operations only one junction admits tunneling at a time, i.e., 
$\delta^I_L=\delta^O_R=0$.
%
Then, we may also derive an
effective feedback master equation for a recursive feedback scheme, where electrons tunneling during a control operation would induce further
control operations -- possibly ad infinitum (at infinite bias and infinite feedback strength).
In this case, the control operations only depend on a single counting field and we may use the decompositions (see appendix~\ref{AeffmasterIIb})
$e^{\kappa_O(\chi_L,\chi_R)} \equiv {\cal P}^O(\chi_L) = {\cal P}^O_N + {\cal P}^O_O e^{+i \chi_L} + {\cal P}^O_I e^{-i \chi_L}$ and
$e^{\kappa_I(\chi_L,\chi_R)} \equiv {\cal P}^I(\chi_R) = {\cal P}^I_N + {\cal P}^I_I e^{-i \chi_R} + {\cal P}^I_O e^{+i \chi_R}$
together with the evident relations ${\cal P}^O_O({\cal F}_L^++{\cal F}_R^+)={\cal P}^I_I({\cal F}_L^-+{\cal F}_R^-)={\cal P}^O_O {\cal P}^I_O={\cal P}^I_I {\cal P}^O_I=\f{0}$ 
(these effectively imply that during the control operations, the transport is unidirectional)
to sum up the infinitely many terms as a von Neumann operator series.
Eventually (see appendix~\ref{AeffmasterIIb}), this results in the effective feedback Liouvillian
\bea\label{EfbIIb}
{\cal L}_{\rm fb}^{IIb}(\chi_L,\chi_R) &=& \sum_{\alpha\in\{L,R\}} \Gamma_\alpha {\cal F}_\alpha^0\nn
&&+ \left({\cal P}^O_N+e^{-i\chi_L} {\cal P}^I_N {\cal P}^O_I\right)\times\nn
&&\;\times \left[\f{1}-{\cal P}^I_O {\cal P}^O_I e^{i(\chi_R-\chi_L)}\right]^{-1}\times\nn
&&\;\times \left(\sum_{\alpha\in\{L,R\}} \Gamma_\alpha {\cal F}_\alpha^+ e^{+i \chi_\alpha}\right)\nn
&&+ \left({\cal P}^I_N+e^{+i\chi_R} {\cal P}^O_N {\cal P}^I_O\right)\times\nn
&&\;\times \left[\f{1}-{\cal P}^O_I {\cal P}^I_O e^{i(\chi_R-\chi_L)}\right]^{-1}\times\nn
&&\;\times\left( \sum_{\alpha\in\{L,R\}} \Gamma_\alpha {\cal F}_\alpha^- e^{-i \chi_\alpha}\right)\,.
\eea

Eqns.~ (\ref{EfbI}),~(\ref{EfbIIa}),~(\ref{EfbIIb}) yield the complete statistics for the current through the SET under the different feedback schemes:
The generating function for its cumulants is in the long term given by the dominant (with $\lambda(0,0)=0$) eigenvalue $\lambda(\chi_L,\chi_R)$ of the effective Liouvillians.
We have checked numerically (see appendixes~\ref{AeffmasterI},~\ref{AeffmasterIIa}, and~\ref{AeffmasterIIb}) that the analytic results from the 
effective master equation coincide for all schemes with an ensemble-average of the 
stochastic Schr\"odinger equation with feedback explicitly included.
In the idealized classical limit, none of the schemes performs work on the system.
However, quantum-mechanically the time-dependent modification of the tunneling rates changes the energy spectrum of the total Hamiltonian 
and thereby performs work~\cite{kim2011a}.

\section{Current and Fluctuations}\label{Scurrent}
We summarize the behavior of the current for finite feedback strengths at reverse infinite bias ($f_L=0$, $f_R=1$), 
zero bias ($f_L=f_R=f$), and infinite bias ($f_L=1$, $f_R=0$) for the different feedback schemes
in table~\ref{Tcurrent}.
\begin{table}[ht]
\begin{tabular}{c|c|c|c}
scheme & $V_{\rm bias}\to-\infty$ &  $V_{\rm bias}=0$ &  $V_{\rm bias}\to+\infty$\\
\hline
I & $-e^{-\delta_{\rm fb}}/2$ & $2 f(1-f) \sinh(\delta_{\rm fb})$ & $e^{\delta_{\rm fb}}/2$\\
IIa & $-1/2$ & $f(1-f)(1-e^{-\delta_{\rm fb}})$ & $1 - e^{-\delta_{\rm fb}}/2$\\
IIb & $-1/2$ & $\frac{f(1-f)(e^{\delta_{\rm fb}}-1)}{2f(1-f)+e^{\delta_{\rm fb}}[1-2f(1-f)]}$ & $e^{\delta_{\rm fb}}/2$
\end{tabular}
\caption{\label{Tcurrent}
Values of the current (in units of $\Gamma_L=\Gamma_R=\Gamma$) under finite but symmetric feedback.
The feedback parameters have been chosen as $\delta^F_L=\delta^E_R=-\delta_{\rm fb}$ and $\delta^F_R=\delta^E_L=\delta_{\rm fb}$
for scheme I and $\delta^I_R=\delta^O_L=\delta_{\rm fb}$ as well as $\delta^I_L=\delta^O_R=0$ for schemes IIa and IIb, respectively.
}
\end{table}
For finite temperatures (where $0 < f < 1$) feedback may induce a current even at zero bias, such that the device acts as a demon
shuffling electrons from one bath to another (here from left to right) using only information on SET occupancy.
This has to be contrasted with feedback ratchets~\cite{cao2009a} where the time-dependent ratchet potential performs work.
The apparent divergence of the current in feedback scheme I for infinite bias and feedback strength is natural as the tunneling rates diverge likewise.
For feedback scheme IIb however, we observe a genuine feedback catastrophe for large bias and large feedback strength leading to a divergent current:
Control operations mutually trigger further control operations with vanishing halting probability, which leads to avalanche-like transport. 
Therefore, in contrast to scheme I, the Fano factor $F=S/\abs{I}$ given by the ratio of noise and current -- as summarized in Table~\ref{Tfano} -- 
also diverges in this region when we let the feedback strength go to infinity.
\begin{table}[ht]
\begin{tabular}{c|c|c}
scheme & $V_{\rm bias}=0$ & $V_{\rm bias}\to+\infty$\\
\hline
I & $\coth(\delta_{\rm fb})/2 + (1-2f)^2 \tanh(\delta_{\rm fb})/2$& $1/2$\\
IIa & $\frac{2f(1-f)+\cosh(\delta_{\rm fb})\left[1-2f(1-f)\right]+\sinh(\delta_{\rm fb})}{(1-2f)^2\left[\cosh(\delta_{\rm fb})-1\right]+\sinh(\delta_{\rm fb})}$ 
& $\frac{4-3 e^{-\delta_{\rm fb}}}{4-2 e^{-\delta_{\rm fb}}}$\\
IIb & $\frac{2 e^{2 \delta_{\rm fb}}-e^{\delta_{\rm fb}}+1}{e^{2\delta_{\rm fb}}-1}$ & $e^{\delta_{\rm fb}}-1/2$
\end{tabular}
\caption{\label{Tfano}
Values of the Fano factor (assuming $\delta_{\rm fb}\ge 0$) for zero and infinite bias with the same parameters as in table~\ref{Tcurrent}, 
the zero-bias Fano factor for scheme IIb has been evaluated for $f=1/2$ only
for brevity.
The reverse infinite bias Fano factor (not shown) is $1/2$ for all schemes independent of feedback.
Divergence of the infinite bias Fano factor for scheme IIb at infinite feedback strength $\delta_{\rm fb}\to\infty$ demonstrates the feedback catastrophe.
Also, at zero feedback the Fano factor should diverge at zero bias as this is the point where conventionally the current would vanish.
}
\end{table}

\begin{figure}[ht]
\includegraphics[width=0.48\textwidth,clip=true]{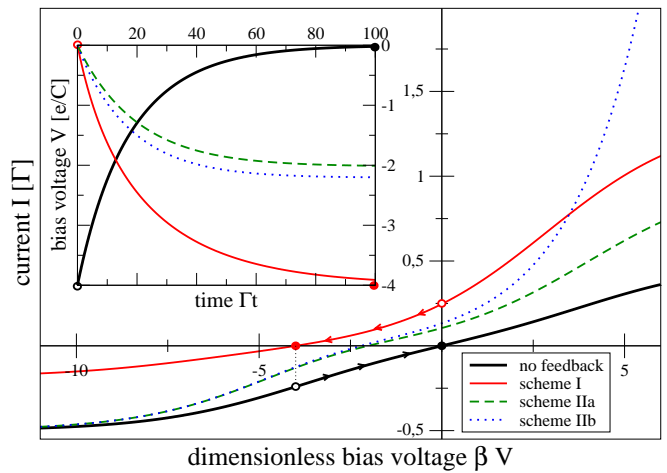}
\caption{\label{Fcurvoltcomp}(Color Online)
Current-voltage characteristics under feedback strength $\delta^E_L=\delta^F_R=-\delta^E_R=-\delta^F_L=1$ in scheme I (thin solid curve, red), maximum feedback
in schemes IIa and IIb (dashed and dotted, respectively) and no feedback (bold black) for symmetric tunneling rates $\Gamma_L=\Gamma=\Gamma_R$ and $\beta\epsilon=2$.
The current may point in the other direction than the voltage leading to a positive power generated by the device.
The inset (for $\beta e/C=1$) shows the mean-field evolution of the voltage from an initial value (empty circles) to the stable fixed points (filled circle) for symmetric capacitances of the two
conductors.
}
\end{figure}

Naturally, there exists a parameter regime where the device transports electrons against an existing 
electrical or thermal gradient, where e.g., electrons are transported from left to right even though $f_L < f_R$, compare also Fig.~\ref{Fcurvoltcomp}.
For scheme I, this effectively implements a Parrondo game (ratchet): Playing two losing strategies (tunneling with the bias) in an alternating 
manner may yield a winning strategy (tunneling against the bias)~\cite{harmer1999a}.
For our model, feedback is necessary to achieve such a current inversion:
Without feedback, the long-term cumulant-generating function for the current 
obeys the analytic relation 
\mbox{$\lambda(0,-\chi) = \lambda\left(0,+\chi- i \ln\left[\frac{(1-f_L) f_R}{f_L (1-f_R)}\right]\right)$},
which when both leads are at the same temperature eventually leads to the fluctuation theorem~\cite{esposito2009a,saito2008a,utsumi2009a}
$\lim\limits_{t\to\infty} \frac{P_n(t)}{P_{-n}(t)} = e^{+ n \beta (\mu_L-\mu_R)}$.
This implies that for a constant Liouvillian (\ref{Emaster}), the current $I=\lim_{t\to\infty} d/dt \sum_n n P_n(t)$ always 
flows from reservoirs with large chemical potential towards the reservoir with the smaller chemical potential.
Also with open-loop feedback, where control operations are applied unconditionally (e.g., in a random or periodic manner), we find that the
current will never flow against the electro-thermal gradient -- as long as no work is performed on the
SET itself (i.e., the SET level $\epsilon$ remains unchanged), see appendix~\ref{Aunconditional}.

For scheme I we find that
the Johnson-Nyquist relation normally relating noise with differential conductance at zero bias
voltage (at similar temperatures left and right) is now shifted to the equilibrium voltage $V^*=\left(-\delta^E_L-\delta^F_R+\delta^E_R+\delta^F_L\right)/\beta$ 
at which the current vanishes $I(V^*)=0$, i.e., more explicitly 
we have $\left.S\right|_{V=V^*} = \frac{2}{\beta} \left.\frac{dI}{dV}\right|_{V=V^*}$.
We obtain a similarly simple modification of the fluctuation theorem for the electron counting statistics under
feedback: 
The cumulant-generating function for the current obeys
\bea
\lambda^I(-\chi) = \lambda^I\left(+\chi-i  \ln\left[\frac{(1-f_L) f_R}{f_L (1-f_R)} e^{\beta V^*}\right]\right)\,,
\eea
which for both leads at the same temperature yields for the counting statistics
$\lim\limits_{t\to\infty} \frac{P_n(t)}{P_{-n}(t)} = e^{n \beta(\mu_L-\mu_R - V^*)}$.
This formula directly demonstrates that the current may be directed against a potential gradient when the feedback parameters and hence $V^*$ are adjusted
accordingly.

For feedback schemes IIa and IIb, the equilibrium voltage $V^*$ can be obtained numerically and depends on the baseline tunneling rates $\Gamma_\alpha$.

\section{Maximum Power at maximum Feedback}\label{Spower}

For schemes IIa and IIb we may take $\delta^I_R=\delta^O_L\to\infty$ and $\delta^I_L=\delta^O_R=0$ to obtain the maximum effect, but to bound the Liouvillian 
for scheme I we constrain ourselves to finite feedback strengths $\delta^E_L=\delta^F_R=+\delta_{\rm fb}$ and $\delta^E_R=\delta^F_L=-\delta_{\rm fb}$.
The power $P=-I V$ generated by the demon can -- when both leads are at the same temperature $\beta$ be 
evaluated as 
$P \beta = -I \beta(\mu_L-\mu_R) = -I(f_L,f_R) \ln\left[\frac{f_L(1-f_R)}{f_R(1-f_L)}\right]$.
The right hand side of this expression can for given $\Gamma_L$ and $\Gamma_R$ be numerically maximized with respect
to the Fermi functions $f_{L/R}\in(0,1)$ yielding the maximum power 
generated by the device, see table~\ref{Tpower}.
\begin{table}[ht]
\begin{tabular}{c|c|c}
scheme & $\Gamma_L=\Gamma_R=\Gamma$ & $\Gamma_{\rm max} \gg \Gamma_{\rm min}$\\
\hline
IIa & $0.0841\; \Gamma k_B T$ & $0.1219\; \Gamma_{\rm max} k_B T$\\
IIb & $0.1892\; \Gamma k_B T$ & $0.1589\; \Gamma_{\rm max} k_B T$ 
\end{tabular}
\caption{\label{Tpower}
Maximum power generated with the device for schemes IIa and IIb at maximum unidirectional feedback for symmetric and highly asymmetric tunneling rates.
}
\end{table}
The maximum power generated with scheme I is unbounded (as the Liouvillian) and asymptotically 
approaches $0.2785\; \Gamma e^{\delta_{\rm fb}} k_B T$ for symmetric tunneling
rates.
Even under idealized conditions, the associated work performed by the demon has to be contrasted with the heat dissipated when
the QPC trajectory's data-points are deleted (Landauer principle).
To perform the continuous monitoring of the SET it is important that the QPC sampling rate $\Delta\tau^{-1}$ is greater than the maximum tunneling rate
$\Delta\tau < \Gamma_{\rm max}^{-1}$ (or $\Delta t \le \Gamma_{\rm max} e^{-\delta^{\rm fb}}$ for scheme I).
This implies that the maximum work per current measurement $W=P_{\rm max} \Delta \tau$ is always smaller than the dissipated Landauer 
heat $Q \ge k_B T \ln 2$ by the demon, such that the results are compatible with the second law.

\section{Feedback Charging Effects}\label{Scharging}

Small leads of finite capacitances will usually be driven out of equilibrium due to transport.
Additional larger reservoirs at thermal equilibrium may however immediately re-enforce equilibrium in the leads solely by scattering
interactions (without electron tunneling).
We may phenomenologically include this effect by making the chemical potentials dependent on
the number of tunneled particles:
Starting at equilibrium, the difference $V$ in chemical potentials between the two leads is classically simply proportional to
the number of tunneled particles $V = e n/C$, where $C$ denotes the total capacitance (similar relations hold for left and right chemical 
potentials in case of asymmetric capacitances~\cite{jackson1998}).
We may numerically solve the resulting mean-field nonlinear (compare appendix~\ref{Anonlinear}) differential equation $\dot{V}=e/C \expval{\dot{n}(t)} = e I(V)/C$
to obtain the dynamical evolution $V(t)$, compare the inset in Fig.~\ref{Fcurvoltcomp}.
After the equilibrium voltage  ($I(V^*)=0$) has been approached, feedback may be stopped and the current will reverse its direction (black trajectory).
With the QPC already in place, it appears reasonable to use it as a detector to clearly discriminate the resulting initial fluctuations from 
equilibrium ones.
This scheme works as an accumulator undergoing charging (feedback) and discharging (no-feedback) cycles, where the total stored energy is
given by $W=C (V^*)^2/(2 e)$.

\section{Experimental implementation}\label{Sexperiment}
With gate voltages, the height of the tunneling barriers may be adjusted such that the timescale
on which electrons tunnel through the QPC can be tuned from ms~\cite{gustavsson2006a} to seconds and even hours.
For periodically varying gate voltages, the frequency corresponds with $100$ MHz to switching times five orders of magnitude smaller in recent
electron pumping experiments~\cite{blumenthal2007a,maire2008a}.
The bandwidth of typical experimental QPC detector devices has been reported in the range of $40$ kHz, with sufficiently larger 
current sampling frequencies of $100$ kHz~\cite{flindt2009a}.
The experimental challenge therefore clearly lies in the necessity of strongly modifying the tunneling rates without performing
work on the system (changing the SET level).
With gate electrodes of sizes below $100$ nm~\cite{maire2008a} it should be possible to keep the energy level of the SET (size about $300$ nm~\cite{gustavsson2006a}) 
approximately constant.

\section{Summary}\label{Ssummary}
To conclude, we have compared several closed-loop feedback schemes implementing Maxwell's demon by means of an effective feedback master equation.
Scheme I used a piecewise constant Liouvillian conditioned on the time-dependent SET occupation, whereas schemes IIa and IIb were conditioned on
its change and used $\delta$-kicks in the tunneling rates.
All schemes are capable of generating a current against a moderate bias -- which is for constant SET level not possible for open-loop schemes -- and may for finite-size 
leads be used to charge a feedback battery.
With Landauers principle, the second law is of course respected by the device.
Schemes I and IIa (with necessarily small feedback strength) should be implementable with present-day technology, whereas the feedback recursion depth appears to be currently 
limited by the QPC sampling frequency.
Schemes with finite recursion depth are however also treatable with the methods in this article.

\section{Acknowledgements}
Financial support by the DFG (grant SCHA 1646/2-1, SFB 910) and stimulating discussions with T.~Novotn\'y and M.~Rinck are gratefully acknowledged.



\appendix
  
\section{Justification of scheme I}\label{AeffmasterI}

Assuming that the dot is empty at time $t$, i.e., $\rho(t)=(1,0)$, its no-measurement time evolution under
feedback scheme I will be governed by $\rho(t+\Delta t) = e^{{\cal L}_E(\chi_L,\chi_R) \Delta t}\rho(t)$, 
whereas for an initially filled dot $\rho(t)=(0,1)$, we will have the evolution  
$\rho(t+\Delta t) = e^{{\cal L}_F(\chi_L,\chi_R) \Delta t}\rho(t)$.
Using projection superoperators on the empty and filled dot states, respectively, both cases
can be incorporated into a single equation
\bea
\rho(t+\Delta t) &=& 
e^{{\cal L}_E(\chi_L,\chi_R) \Delta t}
\left(\begin{array}{cc}
1 & 0\\
0 & 0
\end{array}\right) \rho(t)\nn
&&+
e^{{\cal L}_F(\chi_L,\chi_R) \Delta t}
\left(\begin{array}{cc}
0 & 0\\
0 & 1
\end{array}\right)
\rho(t)\,.
\eea
Expanding the propagators for small $\Delta t$, using that the projection operators add up to the identity, and
solving for the finite-difference $[\rho(t+\Delta t)-\rho(t)]/\Delta t$ yields the effective feedback 
Liouvillian Eq.~(\ref{EfbI}) when $\Delta t \to 0$.
The switching between the different propagators is assumed to be instantaneous, such that during the switching time,
no particles may tunnel. 
\begin{widetext}
Therefore, the original counting fields may be kept, and effectively, the Liouvillian~(\ref{EfbI}) has the first column
from ${\cal L}_E$ and the second column from ${\cal L}_F$
\bea
{\cal L}_{\rm fb}^I(\chi_L,\chi_R) &=& \left(\begin{array}{cc}
-\Gamma_L e^{\delta_L^E} f_L - \Gamma_R e^{\delta_R^E} f_R & +\Gamma_L e^{\delta_L^F} e^{+i\chi_L} (1-f_L) +\Gamma_R e^{\delta_R^F} e^{+i\chi_R} (1-f_R)\\
+\Gamma_L e^{\delta_L^E} e^{-i\chi_L} f_L  +\Gamma_R e^{\delta_R^E} e^{-i\chi_R} f_R &  -\Gamma_L e^{\delta_L^F} (1-f_L) -\Gamma_R e^{\delta_R^F} (1-f_R)
\end{array}
\right)\,,
\eea
which explicitly breaks detailed balance.
\end{widetext}

\begin{figure}[ht]
\begin{center}
\includegraphics[width=0.48\textwidth,clip=true]{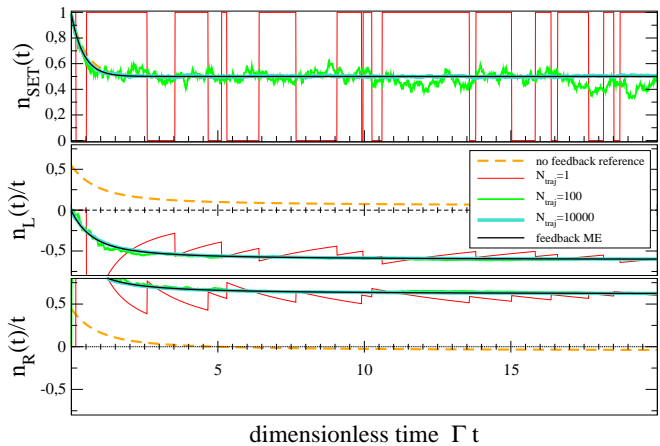}
\end{center}
\caption{\label{FtrajectoriesI}(Color Online)
Comparison of a single (thin red curve with jumps, same realization in all panels) and the average of 100 (medium thickness, green) and 10000 (bold smooth curve, turquoise) 
trajectories with the solution from the effective
feedback master equation [Eq.~(\ref{EfbI}), thin black] for the dot occupation (top), the number of particles on the left (middle), and the number of 
particles on the right (bottom).
The average of the trajectories converges to the effective feedback master equation result.
The reference curve without feedback (dashed orange) may be obtained from Eq.~(\ref{Emastern}) or by using vanishing feedback parameters
and demonstrates that the direction of the current may actually be reversed via sufficiently strong feedback.
Parameters: $\Gamma_L=\Gamma_R\equiv\Gamma$, $f_L=0.45$, $f_R=0.55$, $\delta^E_L=\delta^F_R=1.0$, $\delta^E_R=\delta^F_L=-10.0$, and $\Gamma \Delta t=0.01$.
}
\end{figure}

A single trajectory for this feedback scheme may be generated as follows:
Assuming that e.g., the dot is initially filled, the probability that during a small timestep $\Delta t$ 
the electron jumps out to the left is given by
$P^F_{\rm left}(\Delta t) = \Delta t \Gamma_L e^{\delta^F_L} (1-f_L)$, 
whereas the probability to jump out to the right reads
$P^F_{\rm right}(\Delta t) = \Delta t \Gamma_R e^{\delta^F_R} (1-f_R)$.
The probabilities for electrons jumping from left or right lead into an initially empty dot are obtained similarly and
read
$P^E_{\rm left}(\Delta t) = \Delta t \Gamma_L e^{\delta^E_L} f_L$ and
$P^E_{\rm right}(\Delta t) = \Delta t \Gamma_R e^{\delta^E_R} f_R$.
Using a random number generator, one may with sufficiently small timesteps (such that the jump probabilities are significantly smaller than one)
generate single trajectories for the dot occupation $n_{\rm SET}$, the number of tunneled particles to the left lead $n_L$, and
the number of particles tunneled to the right lead $n_R$.
The ensemble average of many such trajectories may now be compared with the analytic solution of the effective feedback master equation~(\ref{EfbI}),
see Fig.~\ref{FtrajectoriesI}.
It follows that for sufficiently many trajectories the same average observables will be obtained as with the effective feedback 
master equation.


\section{$\delta$-kick propagators}\label{Adeltakick}

There are multiple derivations of delta-kick propagators,
we provide a simple pedestrians derivation based on piecewise constant time dependencies as used in scheme I.
When ${\cal L}(t)$ is a matrix that does not commute with itself at different times, the equation 
$\dot{\rho}(t) = {\cal L}(t) \rho(t)$
is impossible to solve analytically in the general case.
However, when the Liouvillian ${\cal L}(t)$ is piecewise constant, the solution may be readily obtained
by conventionally propagating for a time period where the Liouvillian is constant, and using the 
resulting state as an initial value for the next propagation period with another constant Liouvillian.

Let us therefore assume a constant baseline Liouvillian superimposed with an additional control 
Liouvillian ${\cal L}(t) = {\cal L}_0 + {\cal L}_c \Theta(t-t_c) \Theta(t_c+\tau_c-t)$, 
where the latter is turned on at time $t_c$ and lasts for timespan $\tau_c$ ($\Theta(x)$ denotes the Heaviside step function).
The solution for all times reads in this case
\bea
\rho(t) = \left\{\begin{array}{ccc}
e^{{\cal L}_0 t} \rho_0 & : & t \le t_c\\
e^{({\cal L}_0 + {\cal L}_c)(t-t_c)} e^{{\cal L}_0 t_c} \rho_0 & : & t_c < t < t_c+\tau_c\\
e^{{\cal L}_0 (t-t_c-\tau_c)} e^{({\cal L}_0+{\cal L}_c) \tau_c}e^{{\cal L}_0 t_c} \rho_0 & : & t_c+\tau_c \le t
\end{array}\right.\nonumber\,.
\eea
We now assume that the control Liouvillian scales inversely with the pulse duration ${\cal L}_c = \frac{\kappa_c}{\tau_c}$ with 
$\kappa_c$ being a dimensionless super-operator of Lindblad form.
Letting the pulse duration vanish $\tau_c\to 0$ , we approximate a $\delta$-kick via ${\cal L}(t) = {\cal L}_0 + \kappa_c \delta(t-t_c)$ 
(where $\delta(x)$ represents the Dirac-$\delta$ distribution), such that
the solution reads
\bea
\rho(t) = \left\{\begin{array}{ccc}
e^{{\cal L}_0 t} \rho_0 & : & t \le t_c\\
e^{{\cal L}_0 (t-t_c)} e^{\kappa_c} e^{{\cal L}_0 t_c} \rho_0 & : & t_c \le t
\end{array}\right.\,,
\eea
that is, the constant propagator evolution is simply interrupted by the control operation $e^{\kappa_c}$ at control time $t_c$.

Experimentally, it may be more reasonable to discuss smooth dependencies ${\cal L}(t)={\cal L}_0 + {\cal L}_c(t-t_c)$ with the assumptions
${\cal L}_c(x<-\tau_c/2)={\cal L}_c(x>+\tau_c/2)=\f{0}$ and that $\int {\cal L}_c(t) dt$ is independent of the pulse duration $\tau_c$.
The latter condition also implies that the maximum pulse height must scale inversely with the pulse duration.
In the limit of vanishing pulse duration $\tau_c\to 0$ such a control operation would also converge to a Dirac-$\delta$ distribution, 
such that during control, the baseline Liouvillian ${\cal L}_0$ may be neglected and we 
would have an effective propagator
\bea
\kappa_c \equiv \lim\limits_{\tau_c\to 0} \int\limits_{-\tau_c/2}^{+\tau_c/2} {\cal L}_c (t') dt'\,,
\eea
which must be dimensionless and inherits the Lindblad form from ${\cal L}_c(t)$.


\section{Justification of scheme IIa}\label{AeffmasterIIa}

When ultrashort (in the sense discussed in Appendix~\ref{Adeltakick}) control operations on the 
SET tunneling rates are only to be applied immediately after an electron jumps into or out of the dot, the derivation
of an effective master equation is a bit more involved.
We start from the inverse Fourier transform of the evolution generated by Eq.~(\ref{Emastern})
\bea
\rho^{(n_L,n_R)}(t+\Delta t) &=& {\cal J}^{(n_L, n_R)}(\Delta t) \rho(t)\,,\nn
{\cal J}^{(n_L, n_R)}(\Delta t) &\equiv& \int\limits_{-\pi}^{+\pi} \frac{d^2\chi_\alpha}{4 \pi^2}  e^{{\cal L}(\chi_L,\chi_R) \Delta t 
-i n_L\cdot \chi_L - i n_R \cdot \chi_R}\,,\nn
\eea
where $\rho(t)$ denotes the unconditional dot density matrix at time $t$, and $n_{L/R}$ the number of electrons that have tunneled to left/right reservoirs during the timestep $\Delta t$.
For small $\Delta t$ it suffices to consider single-particle jumps only.
Expanding the propagator for small $\Delta t$ and using the orthonormality relation $\int_{-\pi}^{+\pi} e^{i(n-m)\chi} d\chi=2\pi\delta_{nm}$ we obtain the conditional evolution
equations
\bea
\rho^{(0,0)}(t+\Delta t) &=& \left[\f{1}+\Delta t \left(\Gamma_L {\cal F}_L^0+\Gamma_R {\cal F}_R^0\right)\right]\rho(t)\,,\nn
\rho^{(+1,0)}(t+\Delta t) &=& \Gamma_L \Delta t {\cal F}_L^+ \rho(t)\,,\nn
\rho^{(-1,0)}(t+\Delta t)  &=&  \Gamma_L \Delta t {\cal F}_L^- \rho(t)\,,\nn
\rho^{(0,+1)}(t+\Delta t) &=& \Gamma_R \Delta t {\cal F}_R^+ \rho(t)\,,\nn
\rho^{(0,-1)}(t+\Delta t)  &=&  \Gamma_R \Delta t {\cal F}_R^- \rho(t)\,,
\eea
and the respective probabilities are given by the trace of these operators.
To derive a master equation accounting for the average evolution of observables without feedback one may simply compute the weighted
average $\bar{\rho}(t+\Delta t) = \rho^{(0,0)}+\rho^{(+1,0)}+\rho^{(-1,0)}+\rho^{(0,+1)}+\rho^{(0,-1)}$,
which would after solving for the finite difference scheme $(\bar{\rho}(t+\Delta t)-\rho(t))/\Delta t$ yield the
original Liouvillian in Eq.~(\ref{Emaster}) when $\Delta t\to 0$.
In contrast, with $\delta$-kick feedback operations (compare appendix~\ref{Adeltakick}) e.g.
\bea
\kappa_I &=& \int\limits_{-\tau_c/2}^{+\tau/2} {\cal L}_{I}^{\rm control} (t') dt'\nn
&=& \delta^{I}_L {\cal F}_L(0)+\delta^{I}_R {\cal F}_R(0)\,,
\eea
compare Eq.~(\ref{Emastern})
and similarly for $\kappa_O$, where $\delta^{I/O}_\alpha\equiv\int\limits_{-\tau_c/2}^{+\tau/2} \Gamma^{I/O}_\alpha(t) dt'$ are dimensionless feedback parameters 
(characterizing the product of height and width of the time-dependent control tunneling rates $\Gamma^{I/O}_\alpha(t)$), we have to perform the 
average after applying the matching feedback operation [in (I) or out (O)]
\bea
\bar{\rho}(t+\Delta t) &=& \rho^{(0,0)}+e^{\kappa_O} \rho^{(+1,0)}+e^{\kappa_I} \rho^{(-1,0)}\nn
&&+e^{\kappa_O}\rho^{(0,+1)}+e^{\kappa_I} \rho^{(0,-1)}\,.
\eea
After converting this into a finite-difference equation we would obtain the generator in Eq.~(\ref{EfbIIa}) without counting fields.
Now it is essential that during the control operations, electrons may tunnel through the junctions:
Although the pulse duration is infinitesimally small, the product of pulse height and pulse width remains constant,
such that the tunneling probability is characterized by the feedback strength.
Therefore, counting fields may be re-introduced via a conditional master equation as was done when deriving
Eq.~(\ref{Emastern}) from Eq.~(\ref{Emaster}).
Finally, this yields the form of Eq.~(\ref{EfbIIa}).

For the trajectories we constrain ourselves to opening only one junction at a time for all control operations, e.g., $\delta^I_L=\delta^O_R=0$ 
in order to avoid an unbounded number of particles tunneling during control.
It is straightforward to compute the trajectories as described in appendix~\ref{AeffmasterI}, see Fig.~\ref{FtrajectoriesIIa}.
\begin{figure}[ht]
\begin{center}
\includegraphics[width=0.48\textwidth,clip=true]{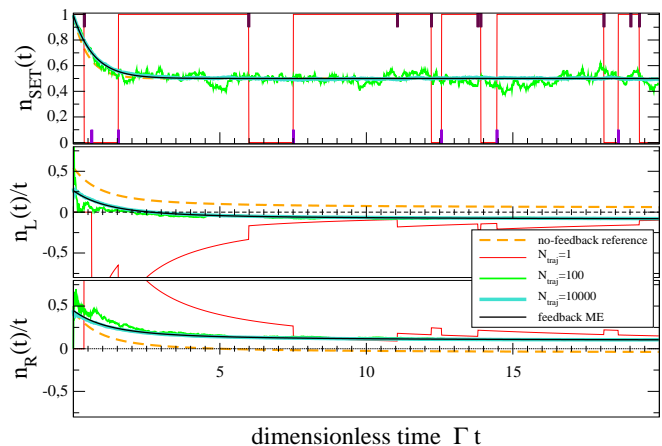}
\end{center}
\caption{\label{FtrajectoriesIIa}(Color Online)
Comparison of a single and the average of 100 and 10000 trajectories with the solution from the effective
feedback master equation~(\ref{EfbIIa}) for the dot occupation (top), the number of particles on the left (middle), and the number of 
particles on the right (bottom).
The average of the trajectories converges to the effective feedback master equation result.
The single trajectory only records the state after all control operations have been performed, i.e., jumps with no net effect are not visible.
Therefore, control operations (bold kinks in top graph) are not always correlated with a net change in the occupation or
the particle number left and right.
The second last top kink for example stands for a out-control operation $e^{\kappa_O}$ that was triggered by an electron jumping to 
the left (an information which is not provided by the QPC) contact.
During the control operation, an electron jumped back from the left contact to the system, such that no net change in $n_{\rm dot}$, 
$n_L$, nor $n_R$ occurred.
Color coding and parameters are as in Fig.~\ref{FtrajectoriesI} where applicable.
Feedback parameters were chosen as $\delta^I_L=\delta^O_R=0$ and $\delta^I_R=\delta^O_L=1.0$.
}
\end{figure}
The computed trajectories do well converge to the results of the effective feedback master equation.
The scheme is non-recursive, i.e., electrons tunneling during control operations do not trigger a further control operation.


\section{Justification of scheme IIb}\label{AeffmasterIIb}

In order to resum the control operations when applied recursively, it is essential that only one junction is opened at a time, 
formally expressed by choosing  $\delta^I_L=\delta^O_R=0$.
This implies that once a control operation has been triggered initially, the transport becomes unidirectional, which 
allows for a simple analytic resummation.
\begin{widetext}
We note that in this case the control operations only depend on a single counting field and we have in their matrix exponential 
the simple decomposition
\bea
e^{\kappa_O(\chi_L,\chi_R)} &\equiv& {\cal P}^O(\chi_L) \equiv {\cal P}^O_N + {\cal P}^O_O e^{+i\chi_L} + {\cal P}^O_I e^{-i\chi_L}
= \left(\begin{array}{cc}
1-(1-e^{-\delta^O_L}) f_L & e^{+i\chi_L}(1-e^{-\delta^O_L})(1-f_L)\\
e^{-i\chi_L} (1-e^{-\delta^O_L}) f_L & 1-(1-e^{-\delta^O_L})(1-f_L)
\end{array}\right)\,,\nn
e^{\kappa_I(\chi_L,\chi_R)} &\equiv& {\cal P}^I(\chi_R) \equiv {\cal P}^I_N + {\cal P}^I_O e^{+i\chi_R} + {\cal P}^I_I e^{-i\chi_R}
= \left(\begin{array}{cc}
1-(1-e^{-\delta^I_R}) f_R & e^{+i\chi_R}(1-e^{-\delta^I_R})(1-f_R)\\
e^{-i\chi_R} (1-e^{-\delta^I_R}) f_R & 1-(1-e^{-\delta^I_R})(1-f_R)
\end{array}\right)\,,
\eea
where the upper index labels the trigger process and the lower indices mark the parts responsible for no particle change (N), 
a particle jumping out of the system (O, possible for $\kappa_I$ only to the right), 
and a particle jumping into the system (I, possible for $\kappa_O$ only from the left).
\end{widetext}
The fact that out of an empty system, no particle may jump out and vice versa for a filled SET system no further particle may jump in is formally reflected 
in the relations ${\cal P}^O_O{\cal F}_R^+={\cal P}^O_O{\cal F}_L^+=\f{0}$ and ${\cal P}^I_I{\cal F}_L^-={\cal P}^I_I{\cal F}_R^-=\f{0}$ as well as
${\cal P}^I_I {\cal P}^O_I={\cal P}^O_O {\cal P}^I_O=\f{0}$.
Therefore, the first-order feedback master equation for scheme IIa in Eq.~(\ref{EfbIIa}) reduces in this case ($\delta^I_L=\delta^O_R=0$) to 
\bea
{\cal L}_{\rm fb}^{(1)} &=& \Gamma_L {\cal F}_L^0 + \Gamma_R {\cal F}_R^0\nn
&&+ ({\cal P}^O_N+{\cal P}^O_I e^{-i\chi_L} )(\Gamma_L e^{+i\chi_L} {\cal F}_L^+ + \Gamma_R e^{+i\chi_R} {\cal F}_R^+)\nn
&&+ ({\cal P}^I_N+{\cal P}^I_O e^{+i\chi_R} )(\Gamma_L e^{-i\chi_L} {\cal F}_L^- + \Gamma_R e^{-i\chi_R} {\cal F}_R^-)\,.\nn
\eea
To generalize this scheme to recursive feedback scheme (IIb) with a potentially infinite recursion depth requires to recursively follow the
scheme ${\cal L}_{\rm fb}^{(i)} \to {\cal L}_{\rm fb}^{(i+1)}$ by performing in each iteration the replacements
\bea
{\cal P}^O_N {\cal A} &\to& {\cal P}^O_N {\cal A}\,,\nn
{\cal P}^O_I {\cal A} &\to& ({\cal P}^I_N+{\cal P}^I_O e^{+i\chi_R}) {\cal P}^O_I {\cal A}\,,\nn
{\cal P}^I_N {\cal A} &\to& {\cal P}^I_N {\cal A}\,,\nn
{\cal P}^I_O {\cal A} &\to& ({\cal P}^O_N+{\cal P}^O_I e^{-i\chi_L}) {\cal P}^I_O {\cal A}
\eea
for arbitrary operators ${\cal A}$ (in this recipe we have also used that ${\cal P}^I_I {\cal P}^O_I={\cal P}^O_O {\cal P}^I_O=\f{0}$), which eventually leads to
\bea
{\cal L}_{\rm fb}^{(\infty)} &=& \Gamma_L {\cal F}_L^0 + \Gamma_R {\cal F}_R^0\nn
&&+({\cal P}^O_N+{\cal P}^I_N {\cal P}^O_I e^{-i\chi_L})\times\nn
&&\;\times \left[\sum_{n=0}^\infty ({\cal P}^I_O {\cal P}^O_I)^n e^{+i n (\chi_R-\chi_L)}\right]\times\nn
&&\;\times (\Gamma_L e^{+i\chi_L} {\cal F}_L^+ + \Gamma_R e^{+i\chi_R} {\cal F}_R^+)\nn
&&+({\cal P}^I_N+{\cal P}^O_N {\cal P}^I_O e^{+i\chi_R})\times\nn
&&\;\times \left[\sum_{n=0}^\infty ({\cal P}^O_I {\cal P}^I_O)^n e^{+i n (\chi_R-\chi_L)}\right]\times\nn
&&\;\times (\Gamma_L e^{-i\chi_L} {\cal F}_L^- + \Gamma_R e^{-i\chi_R} {\cal F}_R^-)\,.
\eea
After summing the von Neumann operator series we finally obtain Eq.~(\ref{EfbIIb}).

\begin{figure}[ht]
\begin{center}
\includegraphics[width=0.48\textwidth,clip=true]{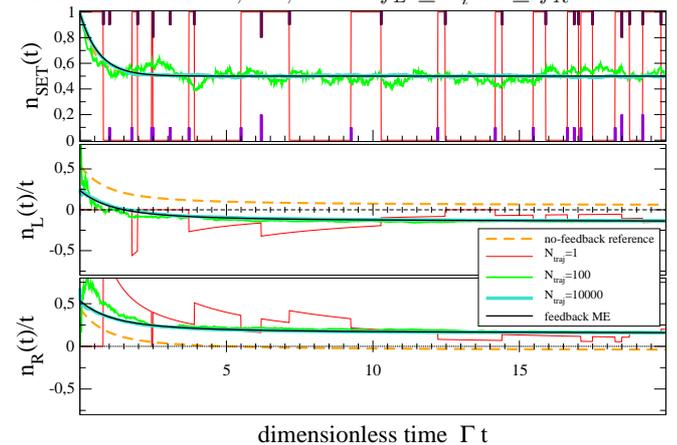}
\end{center}
\caption{\label{FtrajectoriesIIb}(Color Online)
Comparison of a single (only net changes as in Fig.~\ref{FtrajectoriesIIa}) and the average of 100 and 10000 trajectories with the solution from the effective
feedback master equation~(\ref{EfbIIb}) for the dot occupation (top), the number of particles on the left (middle), and the number of 
particles on the right (bottom).
The average of the trajectories converges to the effective feedback master equation result.
Electrons jumping during control operations may now recursively trigger further control operations --
marked by the height of the bold kinks in the top graph.
Color coding and parameters are as in Fig.~\ref{FtrajectoriesIIa}.
}
\end{figure}

Similarly, we may recursively construct the trajectories numerically, see Fig.~\ref{FtrajectoriesIIb}.
The tunneling during mutually calling control operations is only stopped after no particle tunnels.
At infinite bias and infinite feedback strength, the corresponding halting probability vanishes, which leads to
a feedback catastrophe.
Precursors of this are already observed at finite feedback strength and infinite bias in the large current and Fano factor -- compare right columns ins tables~\ref{Tcurrent} and~\ref{Tfano} --
or at infinite feedback strength and finite bias in the current -- compare the dotted curve in Fig.~\ref{Fcurvoltcomp}.


\section{Unconditional control leads to transport with the bias}\label{Aunconditional}

This statement is mainly based on detailed balance, where we assume without loss of generality $f_L < f_R$ at the SET level.

By evaluating the stationary states $\bar{\rho}=(1-\bar{n}^{\rm SET},\bar{n}^{\rm SET})^T$ of Liouvillians ${\cal L}$, ${\cal L}_E$, and ${\cal L}_{F}$ one can see that
their stationary occupation $\bar{n}^{\rm SET}$ is within the transport corridor $f_L \le \bar{n}^{\rm SET} \le f_R$ with the actual position depending 
on the respective tunneling rates.
With any state within the transport corridor as initial condition one can show that the propagators
${\cal P}\in\left\{e^{{\cal L} \Delta t}, e^{{\cal L}_{E} \Delta t}, e^{{\cal L}_{R} \Delta t}, e^{\kappa_I}, e^{\kappa_O}\right\}$ again map to occupations 
$\rho_{i+1}=(1-n^{\rm SET}_{i+1},n^{\rm SET}_{i+1})^T={\cal P} (1-n^{\rm SET}_i,n^{\rm SET}_i)^T={\cal P}\rho_i$ within the corridor, i.e., for 
all $f_L \le n^{\rm SET}_i \le f_R$ we have $f_L \le n^{\rm SET}_{i+1} \le f_R$.
This implies that once any unconditional iterative control scheme has entered the transport corridor (large times), it will not be able to leave it again, regardless of the 
timesteps $\Delta t$, control parameters, baseline tunneling rates, and the order of the operations.
In contrast, with the conditioned feedback scheme one is always outside this corridor with $n^{\rm SET}_i \in\{0,1\}$.
With inserting counting fields at e.g., the right junction ${\cal P}\to {\cal P}(\chi_R)$ one may now calculate the mean 
particle number tunneling during an iteration step
\bea
\Delta n_i = (-i) \partial_{\chi_R} \left.\trace{{\cal P}(\chi_R) (1-n^{\rm SET}_i,n^{\rm SET}_i)^T}\right|_{\chi_R\to 0}\,,
\eea
when the initial state is within the transport corridor ($f_L \le n_i^{\rm SET} \le f_R$).
The outcome is that in this case the electron current always points from right to left when $f_L < f_R$, i.e., for similar temperatures with the bias.

Note also that for unconditional switching between the Liouvillians (scheme I), this is not a Parrondo game anymore, 
since a third possibility of game outcome (no tunneling event at all) is included.


\begin{widetext}
\section{Nonlinear Current-Voltage Characteristics}\label{Anonlinear}

The stationary current (e.g., at the right junction) may either be calculated from the cumulant-generating function or via the relation 
$I=(-i) \trace{\partial_{\chi_R} \left.{\cal L}(\chi_L,\chi_R)\right|_{0,0} \bar\rho}$, 
where ${\cal L}(0,0)\bar\rho=\f{0}$ and reads for scheme I
\bea
I_{\rm fb}^I &=& \frac{\Gamma_L \Gamma_R\left[e^{\delta_L^E+\delta_R^F} f_L (1-f_R)-e^{\delta_L^F + \delta_R^E} (1-f_L) f_R\right]}
{\Gamma_L e^{\delta_L^F} (1-f_L)+\Gamma_R e^{\delta_R^F}(1-f_R) + \Gamma_L e^{\delta_L^E} f_L + \Gamma_R e^{\delta_R^E} f_R}\,,
\eea
where insertion of the Fermi functions $f_\alpha=[e^{\beta_\alpha(\epsilon-\mu_\alpha)}+1]^{-1}$ at similar temperatures
$\beta_L=\beta_R=\beta$ and symmetric chemical potentials $\mu_L=+V/2$ and $\mu_R=-V/2$ yields the full nonlinear current-voltage 
characteristics displayed in Fig.~\ref{Fcurvoltcomp}.
Naturally, the zero-feedback case $\delta_L^E=\delta_R^E=\delta_L^F=\delta_R^F=0$ reproduces the known results.

For feedback schemes IIa and IIb the expressions become a bit lengthy, such that we only give the maximum feedback limit
$\delta^I_L=\delta^O_R=0$ and $\delta^I_R=\delta^O_L\to\infty$ for the current
\bea
I_{\rm fb}^{IIa} &=& \frac{f_L(1-f_R)\left[\Gamma_L^2 (1-f_L)^2 + \Gamma_R^2 f_R^2\right]+\Gamma_L \Gamma_R\left[f_L - f_L^2-f_R^2-f_L f_R-2 f_L^2 f_R^2+2 f_L f_R(f_L+f_R)\right]}
{\Gamma_L\left[1-f_L(1-f_L+1-f_R)\right]+\Gamma_R\left[1-(1-f_R)(f_L+f_R)\right]}\,,\nn
I_{\rm fb}^{IIb} &=& \frac{\Gamma_L^2 f_L (1-f_L) (1-f_R) + \Gamma_R^2 f_L f_R (1-f_R) + \Gamma_L \Gamma_R \left[f_L-f_R^2-f_L^2 f_R+f_L f_R^2\right]}
{\Gamma_L\left[1-f_L(1-f_L+1-f_R)\right]+\Gamma_R\left[1-(1-f_R)(f_L+f_R)\right]}\,.
\eea
Again, insertion of the Fermi functions at similar temperatures and symmetric chemical potentials yields the nonlinear current-voltage characteristics
in Fig.~\ref{Fcurvoltcomp}.
It also becomes evident that at reverse infinite bias ($f_L\to 0$ and $f_R\to 1$), feedback with $\delta^I_L=\delta^O_R=0$ cannot overcome the large bias voltage,
such that the same result as without feedback is obtained.
Inserting these highly nonlinear dependencies in the differential equation $\dot{V} = e I(V)/C$ finally yields the inset of Fig.~\ref{Fcurvoltcomp}.
\end{widetext}


\begin{thebibliography}{9999}

\bibitem{maruyama2009a}
K. Maruyama, F. Nori, and V. Vedral,
Rev. Mod. Phys. {\bf 81}, 1 (2009).

\bibitem{landauer1961a}
R. Landauer,
IBM J.\ Res.\ Dev.\ {\bf 5}, 183 (1961).

\bibitem{datta2008a}
S. Datta, e-print arXiv:0704.1623v1.

\bibitem{esposito2009b}
M. Esposito and K. Lindenberg and C. Van den Broeck,
Europhys.\ Lett.\ {\bf 85}, 60010 (2009).

\bibitem{nazarov2003}
Y. V. Nazarov, {\em Quantum Noise in Mesoscopic Physics}, Kluwer Academic, Dordrecht (2003).

\bibitem{breuer2002}
H.-P. Breuer and F. Petruccione, 
{\em The Theory of Open Quantum Systems},
Oxford University Press (2002).

\bibitem{wiseman2010}
H. M. Wiseman and G. J. Milburn,
{\em Quantum Measurement and Control},
Cambridge University Press (2010).

\bibitem{kiesslich2011}
G. Kiesslich, G. Schaller, C. Emary, and T. Brandes,
to be published by Phys.\ Rev.\ Lett., e-print arXiv:1102.3771.

\bibitem{kim2011a}
S. W. Kim, T. Sagawa, S. De Liberato, and M. Ueda,
Phys.\ Rev.\ Lett.\ {\bf 106}, 070401 (2011).

\bibitem{cao2009a}
F. J. Cao, M. Feito, and H. Touchette,
Physica A 
{\bf 388}, 113 (2009).

\bibitem{harmer1999a}
G. P. Harmer and D. Abbott,
Nature {\bf 402}, 864 (1999).

\bibitem{esposito2009a}
M. Esposito, U. Harbola, and S. Mukamel, 
Rev.\ Mod.\ Phys.\ {\bf 81}, 1665 (2009).

\bibitem{saito2008a}
K. Saito and Y. Utsumi,
Phys.\ Rev.\ B {\bf 78}, 115429 (2008).

\bibitem{utsumi2009a}
Y. Utsumi and K. Saito,
Phys.\ Rev.\ B {\bf 79}, 235311 (2009).

\bibitem{jackson1998}
J. D. Jackson, {\em Classical Electrodynamics}, John Wiley and Sons, New York (1998).

\bibitem{gustavsson2006a}
S. Gustavsson, R. Leturcq, B. Simovic, R. Schleser, T. Ihn, P. Studerus, K. Ensslin, D. C. Driscoll and A. C. Gossard,
Phys.\ Rev.\ Lett.\ {\bf 96}, 076605 (2006).

\bibitem{blumenthal2007a}
M. D. Blumenthal, B. Kaestner, L. Li, S. Giblin, T. J. B. M. Janssen, M. Pepper, D. Anderson, G. Jones, and D. A. Ritchie,
Nature Physics {\bf 3}, 343 (2007).

\bibitem{maire2008a}
N. Maire, F. Hohls, B. Kaestner, K. Pierz, H. W. Schumacher, and R. J. Haug,
Appl.\ Phys.\ Lett.\ {\bf 92}, 082112 (2008).

\bibitem{flindt2009a}
C. Flindt, C. Fricke, F. Hohls, T. Novotny, K. Netocny, T. Brandes, and R. J. Haug,
PNAS {\bf 106}, 10116 (2009).


\end{thebibliography}
\end{document}